\documentclass[%
reprint,
amsmath,amssymb,
aps,
showkeys
]{revtex4-2}

\usepackage{graphicx}
\usepackage{dcolumn}
\usepackage{bm}
\usepackage{hyperref}
\hypersetup{colorlinks,allcolors=blue}

\usepackage{float}
\usepackage[braket, qm]{qcircuit}
\usepackage{graphicx}
\usepackage{multirow}
\usepackage{soul}

\begin{document}

\preprint{APS/123-QED}

\title{NISQ Computers: A Path to Quantum Supremacy}

\author{Muhammad AbuGhanem{$^{1,2,\star}$}}
\author{Hichem Eleuch{$^{3,4,5}$}}
\address{$^{1}$ Faculty of Science, Ain Shams University, Cairo, 11566, Egypt}
\address{$^{2}$ Zewail City of Science, Technology and Innovation, Giza, 12678, Egypt}.
\address{$^{3}$ Department of Applied Physics and Astronomy, University of Sharjah 27272 Sharjah, UAE}
\address{$^{4}$ College of Arts and Sciences, Abu Dhabi University, Abu Dhabi 59911, UAE}
\address{$^{5}$ Institute for Quantum Science and Engineering, Texas A\&M University, College Station, TX 77843, USA}

\email{gaa1nem@gmail.com}

\date{\today}

\begin{abstract}

The quest for quantum advantage, wherein quantum computers surpass the computational capabilities of classical computers executing state-of-the-art algorithms on well-defined tasks, represents a pivotal race in the domain of quantum computing. NISQ (Noisy Intermediate-Scale Quantum) computing has witnessed remarkable advancements, culminating in significant milestones on the journey towards the realization of universal fault-tolerant quantum computers. This transformative turning point, known as quantum supremacy, has been achieved amid a series of breakthroughs, signifying the dawn of the quantum era. Quantum hardware has undergone substantial integration and architectural evolution, contrasting with its nascent stages. 
In this review, we critically examine the quantum supremacy 
experiments conducted thus far, shedding light on their implications and contributions to the evolving landscape of quantum computing. 
Additionally, we endeavor to illuminate a range of cutting-edge proof-of-principle investigations in the realm of applied quantum computing, providing an insightful overview of the current state of applied quantum research and its prospective influence across diverse scientific, industrial, and technological frontiers.

\end{abstract}


\maketitle

\tableofcontents

\section{Introduction}
\label{S:1}

Since its creation, a hundred years ago, quantum mechanics has had something unexpected to offer nearly each single year. The last few decades in particular have seen a dramatic increase in theoretical and practical advances
~\citep{DiVincenzo,MC2,MC3,MC4,MC5,MC6,MC7,MC8,MC9,MC10,MC11,MC12,MC13,MC14,MC15,MC16,MC17,MC18}. 
Among the most compelling frontiers of quantum mechanics is the quest to engineer genuine quantum computers~\citep{DiVincenzo,Nilson,arch02,qc,BA52,Isaac}, machines poised to perform computational tasks far surpassing the capabilities of classical computing counterparts~\citep{nisqQC4,nisqQC6,nisqQC8,nisqQC7,nisqQC9,nisqQC11,nisqQC10}.

At the heart of quantum computing lies the concept of ``computational advantage" or ``quantum supremacy"~\citep{qsuperm1,qsuperm2}. This arises when quantum computers surpass classical computers by executing state-of-the-art algorithms on well-defined tasks~\citep{nisqQC11,nisqQC10}. Notably, only a select few experiments have ventured into the domain of quantum devices with the explicit objective of tackling computational problems that stand beyond the reach of contemporary classical computers~\citep{art19,jiuzhang,Borealis5,Jiuzhang2.0,Zuchongzi}. These groundbreaking experiments centered around the sampling of probability distributions recognized as exponentially challenging for classical computation to simulate effectively. Key milestones include the demonstration of a 53-qubit programmable superconducting processor~\citep{art19} and the realization of a non-programmable photonic system tailored for Gaussian boson sampling, involving 50 squeezed states introduced into a static random 100-mode interferometer~\citep{jiuzhang}. Subsequent iterations of these experiments have witnessed the introduction of additional qubits~\citep{Zuchongzi,Zuchongzi2.1} and the refinement of control over various parameters~\citep{Jiuzhang2.0}. In each case, a crucial benchmark was the comparison of the duration of quantum sampling experiments to the anticipated runtime and scalability of the most advanced classical algorithms, unequivocally situating these quantum platforms within the realm of quantum computational advantage.

These quantum computing milestones stand as remarkable achievements, furnishing researchers and scientists with substantial programmability and control~\citep{art19,jiuzhang,Borealis5,Jiuzhang2.0,Zuchongzi}. However, it is paramount to recognize that these systems have yet to attain fault tolerance~\citep{MC18,faulto1,faulto15,faultoIEEE}. Presently, their operational scope is constrained to executing circuits of modest depth, constrained by notable error rates and decoherence challenges, particularly when managing a relatively low number of qubits (approximately 50 qubits). In spite of commendable efforts to scale up qubit counts, such as IBM's recent unveiling of a quantum computer housing 433 universal and controllable qubits~\citep{ibm433,Ospprey}, and notable progress in advancing photonic quantum computing scalability demonstrated by Xanadu computing~\citep{Borealis}, these systems continue to fall short of the necessary scale required for executing error-correcting codes.

\begin{figure*}
    \centering
    \includegraphics[width=\textwidth]{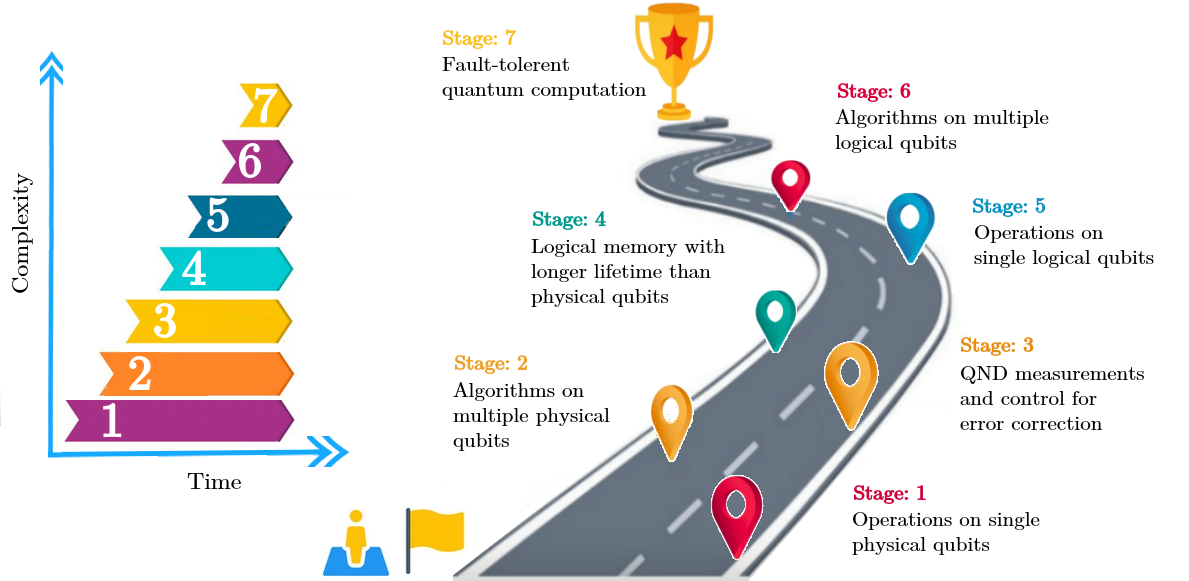}
    \caption{\textbf{The developmental trajectory of quantum information processing.} This development can be conceptualized as a progression through seven distinct stages, each of which relies on and complements the preceding stages while requiring continuous refinement in parallel with the others. Over time, the hope is  to advance from the earlier stages to the more advanced ones, building upon the foundations laid by the preceding phases. Presently, quantum computing platforms are primarily situated between the third stage, involving (QND) Quantum Non-Demolition measurements, and the fourth stage dedicated to logical memory. In this transitional phase, computers are becoming capable of implementing error-correcting codes to construct logical qubits from physical qubits. It's important to note that, with the exception of the last two stages that mark the beginning of the fault-tolerant era, the earlier stages mostly belong to what is known as the NISQ era.
    Additionally, it's crucial to emphasize that progressing from a lower to a higher stage does not negate the ongoing need for improving the techniques required for the lower stages. For instance, enhancing the control of individual physical qubits remains an essential goal throughout the entire developmental timeline, even after logical qubits replace physical ones in computational applications.
}
    \label{fig:7stages}
\end{figure*}

In response to these transformative developments, the new term "NISQ computing" emerged to characterize the prevailing quantum computing era~\citep{jpres18}, differentiating it from the anticipated "fault-tolerant era" that lies in the distant future. It is imperative to underscore that the term "NISQ" primarily conveys a hardware-centric definition and does not inherently imply a temporal dimension. A visual illustration of the developmental trajectory of quantum information processing, with seven distinct stages~\citep{7stages}, is presented in Figure~\ref{fig:7stages}.

As we navigate this quantum odyssey, let us not forget that each step taken in the NISQ era propels us further towards the realization of full fault tolerance and the advancement of increasingly potent quantum devices. While NISQ devices may not yet fully unlock the boundless potential of quantum computing, they provide a fertile ground for exploration and progress. It is imperative to acknowledge that this era may endure for a considerable duration, underscoring the pivotal importance of research during the NISQ era~\citep{jpres18}, given the uncertainties surrounding the timeline for achieving full fault tolerance. Furthermore, NISQ technology holds substantial promise, equipping us with innovative tools to address a multitude of complex challenges~\citep{4Challanges,Challanges}. Moreover, the knowledge and expertise cultivated during the NISQ era are destined to play a pivotal role in propelling us towards the fault-tolerant era, where the true potential of quantum computing awaits realization.

Within this broader context, this review assumes significant relevance, as it aims to provide an in-depth examination of recent 
quantum supremacy experiments and their far-reaching implications. By dissecting the intricacies and consequences of these groundbreaking achievements, we seek to foster a deeper comprehension of the present landscape of quantum computing and its transformative potential. 

\section{Primary Holders of Quantum Computing Patents}

\begin{figure}
    \centering
    \includegraphics[width=0.5\textwidth]{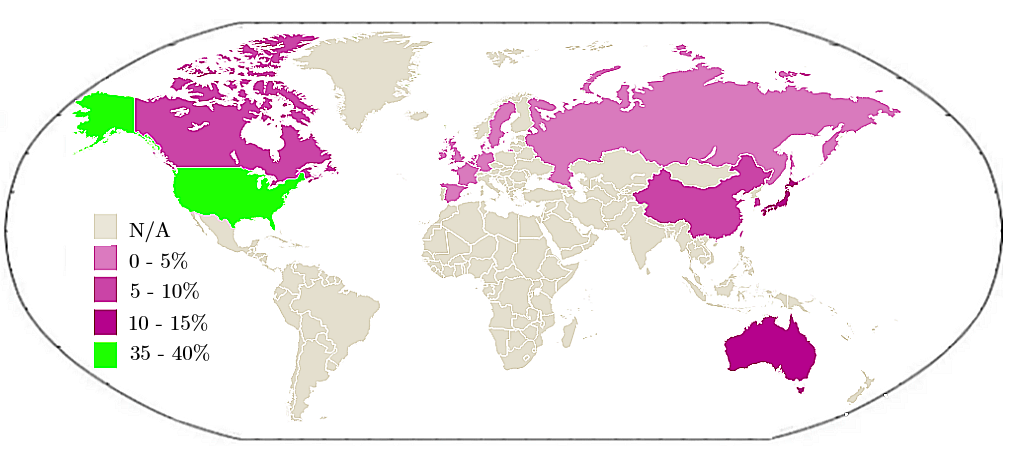}
    \caption{Proportion of patent applications in the field of 2G quantum simulation for which a patent was granted, highlighting the United States' significant role with over 37\% of International Patent Families securing IP rights. It also notes other regions' substantial but relatively lower proportions, including Japan (14\%), Australia (12\%), China (9\%), and Europe (8\%, based on granted European patents). Data sourced from~\citep{EPO23sim}.}
    \label{fig:MAP}
\end{figure}

According to the 2023 report by the European Patent Office (EPO)~\citep{EPO23sim}, a significant metric for assessing the effectiveness and strategic direction of patent filing strategies--in the realm of second-generation quantum computing (2GQC)--is the ratio of granted intellectual property (IP) rights in specific countries or regions. Figure~\ref{fig:MAP} illustrates the portion of ``International Patent Families" within this domain for which IP rights were granted in particular jurisdictions. Notably, over 37\% of these International Patent Families secured at least one IP right in the United States, underscoring its pivotal role in the field of quantum simulation. In comparison, other regions also exhibit substantial proportions, though notably lower than that of the United States. Regions in this regard include Japan (14\%), Australia (12\%), China (9\%), and Europe (8\%, based on granted European patents).

Table~\ref{patents} presents a compilation of leading holders of quantum computing patents, within the past two decades according to the United States (USPTO) and Europe (EPO), as reported in~\citep{USEPOT}. Among the top ten patent holders who have secured more than 25 patents in quantum computing, prominent USA-based multinational enterprises concentrating on both hardware and software platforms (such as IBM, Google, Microsoft, Intel) emerge. Additionally, we identify Northrop Grumman, a U.S. defense technology company, as well as Rigetti, a California-based venture-backed firm that was established in 2013 with a focus on delivering scalable quantum processor technology utilizing superconducting chips. Further contributors to this list include Honeywell International and the USA Government. Turning to the international arena, we encounter D-Wave Systems from Canada, which is particularly dedicated to specialized quantum annealing computers, and Toshiba from Japan. In the realm of academia, noteworthy institutions with substantial portfolios in quantum computing encompass MIT (located in the United States), Oxford (in the United Kingdom), Yale (in the United States), Harvard (in the United States), Caltech (in the United States), Stanford (in the United States), the University of Maryland (in the United States), the University of Wisconsin (in the United States), and the Technical University of Delft in the Netherlands.

\begin{table*}
\centering
\caption{\label{patents} Top primary holders of quantum computing patents in the USPTO and EPO~\citep{USEPOT}. 
}
\begin{ruledtabular}
\begin{tabular}{c|cccc| ccc}
\textbf{No.} & \textbf{Holder} & \textbf{Country} & \textbf{Patents} &\textbf{No.} & \textbf{Holder} &\textbf{Country}&  \textbf{Patents} \\
\hline
1 & IBM & USA & 254 & 29  & STMicroelectronics & Switzerland & 9  \\
2 & D-Wave Systems & Canada & 183 &30 &  Harvard College & USA & 8 \\
3 & Northrop Grumman & USA & 120 &31 &  Magiq Tech Inc & USA & 8 \\
4 & Microsoft Corp & USA & 111 &32 &  California Inst of Tech & USA & 7 \\
5 & Alphabet Inc & USA & 59 &33 &  HP Inc & USA & 7 \\
6 & Rigetti \& Co Inc & USA & 53 &34 &  NEC Corp & Japan & 7 \\
7 & Toshiba Corp & Japan & 37 &35 & Stanford University & USA & 7\\
8 & Intel Corp & USA & 32 &36 &  University Sys of Maryland & USA & 7 \\
9 & Honeywell Int Inc & USA & 26 &37 & University Wisconsin Warf & USA & 7  \\
10 & US Government & USA & 26 &38 &   Kyndryl Inc & USA & 6 \\
11 & HP Enterprise & USA & 23 &39 &  Mitre Corp & USA & 6 \\
12 & Newsouth Innovations & Australia & 22 &40 &  Parallel Investment & USA & 6 \\
13 & Mass Inst of Tech MIT & USA & 20 &41 &QC Ware Corp & USA & 6  \\
14 & Equal1labs Inc & USA & 16 &42 &  Technische Universiteit Delft & Netherlands & 6 \\
15 & Hitachi Ltd & Japan & 15 &43 &  University Johns Hopkins & USA & 6 \\
16 & Japan Science \& Tech Ag & Japan & 15 &44 & Wells Fargo Bank & USA & 6   \\
17 & 1QB Information Tech & Canada & 14 &45 &  Corning Corp & USA & 5\\
18 & Accenture Public Ltd & Ireland & 12 &46 &  Dartmouth College & USA & 5 \\
19 & IonQ Inc & USA & 12  &47 &  Lockheed Martin Corp & USA &  5 \\
20 & Nokia Corp & Finland & 12 &48& Phoenix Co of Chicago & USA & 5\\
21 & Bank of America & USA & 11 &49 &  Quantum Valley Invest & Canada & 5 \\
22 & Element Six SA & Luxembourg &11&50 &  Samsung Electronics & South Korea & 5 \\
23 & Gov. of Abu Dhabi & UAE & 11 &51 &  University California & USA & 5 \\
24 & University Oxford & UK & 11 &52 &  Bull SA & France & 4 \\
25 & Yale University & USA & 10 &53 &  DWSI Holdings Inc & USA & 4 \\
26 & Commissariat Atomique & France & 9 &54 &  IMEC & Belgium & 4 \\
27 & Raytheon Tech Corp & USA & 9 &55 &  Korea Elect. Res Inst & South Korea & 4\\
28 & SEEQC Inc & USA & 9 &56 &  Quintessencelabs Pty & Australia & 4  \\
\end{tabular}
\end{ruledtabular}
\end{table*}

\section{Quantum Supremacy}

In the pursuit of achieving quantum supremacy~\citep{qsuperm1,qsuperm2}, significant efforts have been dedicated to developing quantum computers capable of performing tasks previously deemed impossible. Researchers have made notable strides in demonstrating quantum advantage using both cryogenic and photonic-based approaches. 
A number of experiments have harnessed quantum devices to perform computational tasks that exceed the capabilities of today's classical computers, as documented in various studies~\citep{art19,jiuzhang,Borealis5,Jiuzhang2.0,Zuchongzi}. In all of these experiments, the computational challenges revolved around sampling from probability distributions that are widely acknowledged to be exceedingly difficult for classical computers to simulate efficiently.

As previously mentioned, the concept of quantum supremacy signifies the point at which a quantum computer achieves tasks previously considered beyond the reach of classical computing systems.  Accordingly, two distinct quantum computing paradigms employed for this purpose are Boson Sampling and Gaussian Boson Sampling. Both Boson Sampling (BS)~\citep{BosonS} and Gaussian Boson Sampling (GBS)~\citep{GBS} paradigms exploit the principles of quantum interference among indistinguishable particles, typically bosons, to perform specialized computational tasks.

\subsection{Boson Sampling}

Boson Sampling has emerged as a valuable tool for exploring the quantum advantages over classical computing, particularly because it does not necessitate universal control over the quantum system. This characteristic aligns with the capabilities of existing photonic experimental platforms. Boson Sampling~\citep{BosonS} has captivated both theorists and experimentalists by showcasing the superior computational power of quantum systems and challenging the Extended Church-Turing (ECT) theorem without relying on the full capabilities of a universal quantum computer. 
In Boson Sampling, identical photons traverse a network of optical elements, generating complex interference patterns, and the output distribution encodes information about the underlying quantum states and network parameters. The goal is to sample from this distribution, a task believed to be computationally challenging for classical computers, thus showcasing quantum computational superiority~\citep{BosonS}. Nevertheless, the original Boson Sampling approach encounters significant challenges when it comes to scalability and experimental implementation, chiefly due to the demanding requirement of a reliable source of numerous indistinguishable photons.

\subsection{Gaussian Boson Sampling}

In response to these experimental constraints, a novel approach known as Gaussian Boson Sampling emerged in 2017~\citep{GBS}. GBS tackles a classically hard-to-solve problem by utilizing squeezed states as a non-classical resource. This innovative approach addresses questions regarding the complexity of sampling from a general squeezed state. In GBS, a novel expression has been derived, connecting the probability of measuring a specific output pattern of photons from a general Gaussian state to the \textit{hafnian} matrix function~\citep{GBS}. Gaussian Boson Sampling extends this concept to continuous-variable quantum states and operations, utilizing Gaussian states and transformations. This approach maintains the fundamental principles of interference while dealing with continuous variables like position and momentum.  

Both Boson Sampling and Gaussian Boson Sampling represent key quantum supremacy experiments, highlighting the potential computational advantages of quantum systems in specific, though challenging, computational scenarios. 
For a more in-depth study of Gaussian Boson Sampling (GBS), readers are encouraged to refer to~\citep{GaussianBS2}, and for more understanding of GBS applications, a wealth of information can be found in references~\citep{Borealis3,GBSAPPL13,GBSAPPL12,GBSAPPL14,GBSAPPL16,GBSAPPL15,GBSAPPL18,GBSAPPL17}.

\section{Attaining Quantum Supremacy}

\subsection{Google's Quantum AI}

\subsubsection{Overview}

Random quantum circuits (RQCs) are widely recognized for their complexity, posing significant challenges for classical simulation. The utilization of RQCs played a pivotal role in the pursuit of achieving quantum supremacy~\citep{qsuperm1,qsuperm2}. 
In a notable experiment~\citep{art19}, a programmable quantum processor comprising 54 superconducting qubits performed computations within a Hilbert space of size $2^{53}$. 
The comparative performance of this task on \textit{Summit}, currently the world's most powerful supercomputer, was initially estimated to require approximately 10,000 years for completion. However, the same computation was achieved in a mere 200 seconds using the Sycamore quantum processor.

Shortly after the publication, a discussion ensued regarding the possibility of overestimating the time needed to solve the identical problem on a supercomputer. Within this debate, a significant concern emerged. It is the question, whether reliable methods exist to accurately assess the true capabilities of quantum computing~\citep{depate1,depate2,depate3}. 
Subsequently, it was suggested that an equivalent task could be executed with high fidelity on a classical computer, specifically the \textit{Summit} supercomputer at Oak Ridge National Laboratories, within a matter of days~\citep{depate1}. Additionally, an alternative classical simulation algorithm based on tensor networks was introduced in~\citep{depate2} capable of performing the task in less than 20 days.

Furthermore, an additional tensor network approach was proposed in~\citep{depate3} to address the classical simulation of Sycamore's quantum circuits. This computational task, when executed on a cluster equipped with 512 GPUs, was successfully completed within a span of 15 hours. These findings collectively underscore the immense computational challenges posed by RQCs and the significant strides made in both quantum and classical computing to tackle them.

\subsubsection{Google's Sycamore quantum processor}

\begin{figure}
    \centering
    \includegraphics[width=0.45\textwidth]{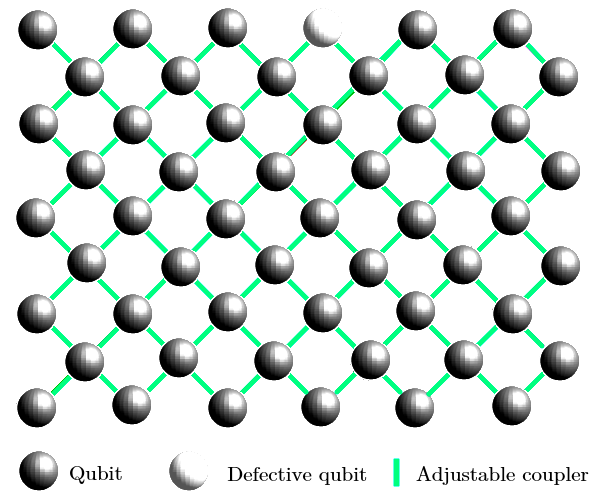}
    \caption{Google's Sycamore quantum processor layout. This processor composed of a rectangular-array of $54$ programmable Superconducting transmon qubits. Each qubit linked to the four nearest neighbors via couplers. As one qubit did not operate perfectly during the supremacy experiment, the quantum processor has $53$-qubits and $86$-couplers.}
    \label{fig:Sycamore}
\end{figure}

The  Google's Sycamore quantum processor~\citep{art19} is composed of a two-dimensional array consisting of 54 programmable Superconducting transmon qubits~\citep{koch07}. Each qubit is configured to have tunable connections with its four nearest neighbors, forming a rectangular lattice as depicted in Figure~\ref{fig:Sycamore}.

\begin{figure*}
    \centering
    \includegraphics[width=\textwidth]{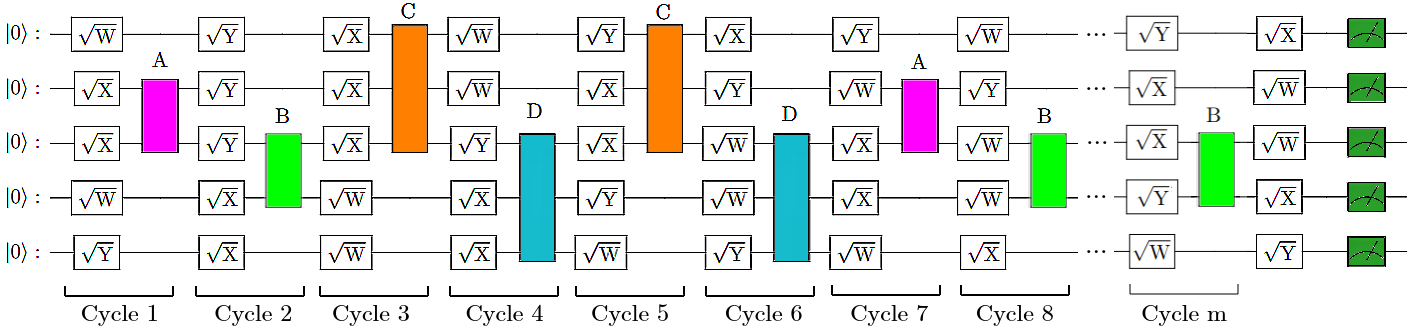}
    \caption{A $m$-cycle circuit schematic diagram of the Sycamore 53-qubit RQCs.
Each cycle joining a layer of random single-qubit gates (chosen at random from the gate set $\{\sqrt{X},\, \sqrt{Y},\, \sqrt{W} \}$), 
followed by a layer of 2-qubit gates, labeled A, B, C, or D.
Layers in longer circuits repeat in the order A; B; C; D -- C; D; A; B. It is worth noting that single-qubit gates do not repeat sequentially, also there is an additional layer of one-qubit gates prior measurements.}
    \label{fig:RQCs}
\end{figure*}

In pursuit of achieving a state of quantum supremacy~\citep{qsuperm1,qsuperm2}, the Sycamore processor introduced significant technological advancements, as detailed in~\citep{art19}. These innovations enabled the execution of random quantum circuits on the 54-qubit Sycamore quantum device, introducing a novel class of random quantum circuits. These circuits are characterized by alternating layers of single-qubit and two-qubit gates, collectively forming a ``cycle." Each random circuit comprises $m$ such cycles, with a ``cycle" defined as a combination of a layer of single-qubit gates followed by a layer of two-qubit gates, concluding with another layer of one-qubit gates before the final measurement step, as illustrated in Figure~\ref{fig:RQCs}.

The Sycamore processor is notable for its capability to execute high-fidelity one-qubit and two-qubit quantum gates. These gates are not only performed individually but also simultaneously across multiple qubits during practical computations~\citep{art19}.

The Sycamore quantum computer employs transmon qubits~\citep{koch07}, which can be regarded as nonlinear superconducting resonators operating at frequencies between 5 and 7 GHz. Each qubit is encoded using the two lowest quantum eigenstates of the resonant circuit. Furthermore, each transmon qubit is equipped with two controls: a microwave drive for qubit excitation and a magnetic flux control for tuning its frequency. For qubit state readout, each qubit is associated with a linear resonator~\citep{wall04}. Additionally, each qubit is interconnected with its neighboring qubits through adjustable couplers~\citep{chen14,yan18}, facilitating dynamic tuning of qubit-qubit coupling, ranging from fully disengaged to 40 MHz. During the quantum supremacy experiment, due to one qubit's malfunction, the device operated with 53 qubits and 86 couplers.

Google Quantum AI and its collaborators unveiled a demonstration that transformed quantum computing from a theoretical concept into a practical reality~\citep{art19}. While Google Quantum AI had not yet achieved the ability to create physical qubits with the desired longevity, it did manage to develop qubits that remained stable long enough to outperform the world's most powerful supercomputer in certain calculations~\citep{art19}. This transition ushered us into the era of Noisy, Intermediate Scale Quantum computing, marking a pivotal shift away from pure classical computing~\citep{jpres18}. A comprehensive characterization of Google's Sycamore quantum processor is presented in~\citep{AGSycam1,AGSycam2}.

\subsection{The Zuchongzhi Quantum Computers}

\subsubsection{Overview}

In another investigation, researchers delved into the capabilities of the Zuchongzi 2.0 quantum computing system, which comprises 56 superconducting qubits~\citep{Zuchongzi}. Their study underscored Zuchongzi's remarkable efficiency in executing complex sampling tasks. Notably, Zuchongzi achieved a sampling task in approximately 1.2 hours, a feat that would require the \textit{Summit} supercomputer, the most powerful classical computer, a minimum of 8 years to complete. Building upon these achievements, the researchers subsequently introduced an enhanced iteration known as Zuchongzhi 2.1, featuring 66 qubits and delivering substantial improvements in quantum advantage compared to its predecessor. These results illuminate the quantum information processing prowess of Zuchongzhi, marking a significant stride in quantum computing hardware and its computational superiority over classical counterparts.

\subsubsection{Zuchongzhi 2.0}

Zuchongzhi 2.0 is a novel two-dimensional programmable superconducting quantum processor comprising 66 functional qubits and 110 couplers~\citep{Zuchongzi}. To evaluate its performance, researchers conducted a benchmarking task involving random quantum circuit sampling, utilizing 56 qubits and 20 cycles. The computational cost of classically simulating this task is estimated to be 2 to 3 orders of magnitude higher compared to prior work with the 53-qubit Sycamore processor~\citep{art19}. Remarkably, Zuchongzhi completed the sampling task in approximately 1.2 hours, a task that would require the most powerful supercomputer at least 8 years to accomplish~\citep{Zuchongzi}.

Zuchongzhi's achievements extend to high-fidelity single-qubit gate operations (averaging 99.86\% fidelity), two-qubit gate operations (averaging 99.41\% fidelity), and high-fidelity readout processes (averaging 95.48\% fidelity). These milestones firmly establish a quantum computational advantage that is unattainable through classical computation within a reasonable timeframe.

\subsubsection{Zuchongzhi 2.1}

In February 2022, a significant milestone in quantum supremacy was reached with the introduction of Zuchongzhi 2.1, a two-dimensional superconducting quantum processor featuring 66 qubits~\citep{Zuchongzi2.1}. Compared to its predecessor, Zuchongzhi 2.0~\citep{Zuchongzi}, Zuchongzhi 2.1~\citep{Zuchongzi2.1} showcases remarkable improvements in readout performance, achieving an average fidelity of 97.74\%.

This enhanced quantum processor empowers the execution of larger-scale random quantum circuit sampling, accommodating a system scale of up to 60 qubits and 24 cycles. The complexity of the achieved sampling task surpasses previous benchmarks, being approximately six orders of magnitude more challenging than that tackled by Google's Sycamore quantum processor~\citep{art19} and three orders of magnitude more complex than the task performed on Zuchongzhi 2.0~\citep{Zuchongzi}.

It was estimated that, state-of-the-art classical algorithms and supercomputers would require approximately 48,000 years to simulate the random circuit sampling experiment, while Zuchongzhi 2.1 accomplishes it in just around 4.2 hours. This substantial reduction in time underscores the significant quantum computational advantage offered by Zuchongzhi 2.1~\citep{Zuchongzi2.1}.

Table~\ref{t:sycamZuchongzhi} presents a comprehensive comparison of system characteristics (mean values) between the Zuchongzhi 2.0 and Zuchongzhi 2.1 superconducting quantum computers. Furthermore, Table~\ref{t:sycamZuchongzhi} provides a performance evaluation of tensor network algorithm runtimes for various circuits executed on the \textit{Summit} supercomputer. It also includes estimates for the computational resources required for classical simulations of the random quantum circuit sampling experiment conducted on the Sycamore quantum processor~\citep{art19}, Zuchongzhi 2.0~\citep{Zuchongzi}, and Zuchongzhi 2.1~\citep{Zuchongzi2.1} quantum processors.

The findings reported in references~\citep{Zuchongzi,Zuchongzi2.1} underscore the remarkable capabilities of Zuchongzhi in quantum information processing, marking a substantial leap forward in quantum computing hardware and its computational superiority when compared to classical counterparts.

\begin{table*}
\caption{\label{t:sycamZuchongzhi} Performance comparison of tensor network algorithm runtimes for various circuits on \textit{Summit} supercomputer. With estimated classical simulation consumption for the random quantum circuit sampling experiment on Sycamore, Zuchongzhi 2.0, and Zuchongzhi 2.1 quantum processors is included~\citep{Zuchongzi2.1}.}
\centering 
\begin{ruledtabular}
\begin{tabular}{lc cc c}
\multirow{2}{*}{Parameters}  & \multicolumn{4}{c}{Google's Sycamore Quantum AI vs Zuchongzhi Quantum Processors} \\ 
\cline{2-5}
 &Sycamore~\citep{art19} &Zuchongzhi 2.0~\citep{Zuchongzi} &Zuchongzhi 2.1~\citep{Zuchongzi2.1} &Zuchongzhi 2.1~\citep{Zuchongzi2.1} \\
\hline 
Quantum Circuit                    &53-qubit 20-cycle    &56-qubit 20-cycle &60-qubit 22-cycle &60-qubit 24-cycle\\
Estimated Runtime on \textit{Summit} & \st{10,000 years} 15.9 days\footnotemark[1]  & 8.2 years &$4,800$ years &$48,000$ years \\ 
Runtime on Quantum Computer & 3 minutes\footnotemark[2]  & 72 minutes& 60 minutes & 252 minutes \\
Fidelity                    &0.224\%       &0.0662\%  &0.0758\% &0.0366\%\\
\end{tabular}
\end{ruledtabular}
\footnotetext[1]{Google's estimations projected that the task would demand an astounding 10,000 years for \textit{Summit}, the most powerful supercomputer. Subsequently, IBM provided a contrasting perspective, asserting that \textit{Summit} could accomplish the same task in a matter of days~\citep{depate1,Zuchongzi2.1}.}
\footnotetext[2]{As reported by Google's Quantum AI and Collaborators in~\citep{art19}, while the time estimated for performing the task in~\citep{Zuchongzi2.1} was 600 seconds.}
\end{table*}

\subsection{The Photonic Quantum Computers, Jiuzhang}

\subsubsection{Overview}

Researchers at the University of Science and Technology in China (USTC) have accomplished a remarkable feat in the realm of quantum computing by achieving quantum computational advantage using novel photonic quantum computers. These cutting-edge quantum computing systems harness the power of photons, or particles of light, to perform calculations that surpass the capabilities of traditional computers. 
In December 2020, the photonic quantum computer "Jiuzhang" achieved quantum advantage~\citep{jiuzhang}. However, it lacked programmability due to its reliance on fixed optical components. Whereas, 
in 2021, researchers in USTC reported on the improved photonic quantum computer "Jiuzhang 2.0," capable of large-scale Gaussian boson sampling~\citep{Jiuzhang2.0}. 
This successful demonstration of quantum supremacy by photonic quantum computers represents a significant leap forward in the field of quantum computing.

\subsubsection{Jiuzhang}

\begin{figure}
    \centering
    \includegraphics[width=0.47\textwidth]{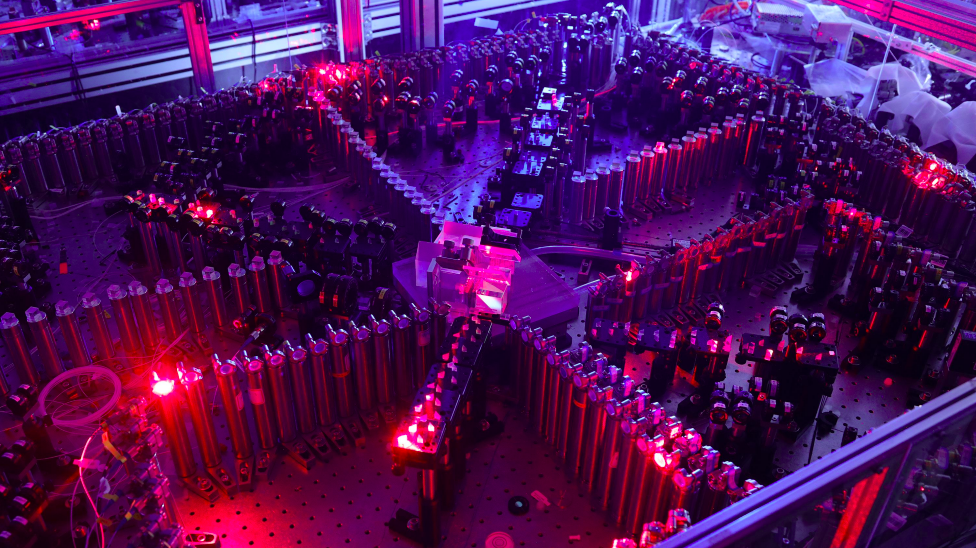}
    \caption{
The photonics-based quantum computing device developed by USTC, known as Jiuzhang, operates by intricately controlling light through an array of optical devices. 
A photographic representation of the Jiuzhang photonic network reveals the experimental configuration, which is situated on an optical table spanning approximately three square meters. Within this setup, 25 Two-Mode Squeezed States (TMSSs) are introduced into the photonic network, resulting in the collection of 25 phase-locked light signals. To further elaborate, the output modes of the Jiuzhang photonic network are meticulously separated into 100 distinct spatial modes through the utilization of mini-mirrors and Polarizing Beam Splitters (PBSs). 
This achievement marks the second quantum computer claiming to attain quantum computational advantages, following Google's Sycamore quantum processor. Reproduced under Creative Commons license CC BY from~\citep{jiuzhang}.}
    \label{fig:Jiuzhang}
\end{figure}

These photonic quantum computers, known as Jiuzhang~\citep{jiuzhang} and Jiuzhang 2.0~\citep{Jiuzhang2.0}, mark significant milestones in the advancement of photonic quantum computing technology. Jiuzhang achieved the distinction of being the first to attain quantum supremacy, while Jiuzhang 2.0 has made substantial progress in terms of programmability and speed, particularly in executing large-scale Gaussian boson sampling tasks~\citep{BosonS,GBS,GBS00,GBS1}. As research and development in the field of quantum computing continue to advance, further enhancements and refinements in both systems are expected to shape the landscape of quantum information processing.

\textit{H. -S. Zhong et al.} conducted an experiment~\citep{jiuzhang} wherein they introduced 50 identical single-mode squeezed states into a 100-mode ultralow-loss interferometer, and subsequently, they captured the outcomes using 100 highly efficient single-photon detectors. By achieving up to 76-photon coincidences, corresponding to a state space dimension of approximately $10^{30}$, they ascertained a sampling rate approximately $10^{14}$ times faster compared to contemporary classical simulation techniques and supercomputing systems.

To provide context, they also computed the time required by two supercomputers to accomplish the same GBS task. \textit{H. -S. Zhong et al.} projected that the \textit{TaihuLight} supercomputer would necessitate around $8\times 10^{16}$ seconds (which is 2.5 billion years), while \textit{Fugaku} would take $2\times 10^{16}$ seconds (which is 0.6 billion years)~\citep{jiuzhang}. The light-based quantum computer Jiuzhang is shown in Figure~\ref{fig:Jiuzhang}.

\subsubsection{Jiuzhang 2.0}

\begin{figure*}
    \centering
    \includegraphics[width=\textwidth]{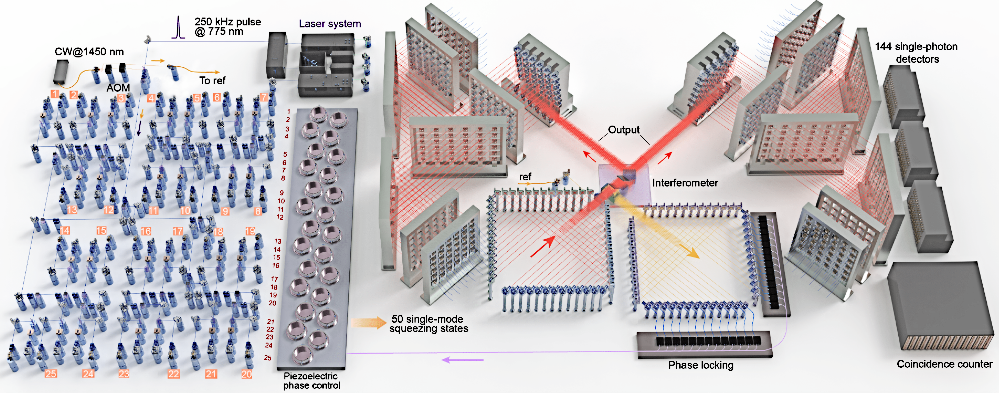}
    \caption{\textbf{The Jiuzhang light-based quantum computer.}
In the upper-left quadrant of its configuration, a high-intensity pulse laser emitting light with a wavelength of 775 nm is utilized to stimulate 25 sources of Two-Mode Squeezed States (TMSS), indicated by the orange label in the left section.
Simultaneously, a continuous-wave laser emitting light at 1450 nm is directed to co-propagate alongside the aforementioned 25 TMSS sources. The resulting output is two-mode squeezed light at 1550 nm, which is then guided into a single-mode fiber known for its temperature insensitivity.   
Notably, a 5-meter segment of this fiber is wrapped around a piezo-electric cylinder, allowing for precise control over the source phase, situated in the central region of the setup. Transitioning to the center-right section, an optical arrangement involving collimators and mirrors facilitates the injection of the 25 TMSSs into a photonic network. In this context, 25 light beams corresponding to these TMSSs (depicted in yellow) with a wavelength of 1450 nm and an intensity power of approximately 0.5 $\mu$W are employed for the purpose of phase synchronization.
The output modes resulting from this setup, totaling 144 modes, are divided into four segments through arrays of adjustable periscopes and mirrors. Ultimately, these output modes are subjected to detection using 144 superconducting nanowire single-photon detectors and subsequently processed through a 144-channel ultra-fast electronics coincidence unit. 
Reproduced under Creative Commons license CC BY from~\citep{Jiuzhang2.0}.}
    \label{fig:JiuzhangQC}
\end{figure*}

In 2021, researchers from the USTC introduced an advanced photonic quantum computer called "Jiuzhang 2.0," designed specifically for large-scale Gaussian boson sampling (GBS)~\citep{Jiuzhang2.0}. This quantum computer utilized a 144-mode photonic circuit, achieving up to 113 photon detection events. A notable achievement was the development of a novel and scalable quantum light source based on stimulated emission of squeezed photons, which exhibited both near-unity purity and efficiency.

The remarkable photonic quantum computing capabilities of Jiuzhang 2.0 correspond to an impressive Hilbert space dimension of approximately $10^{43}$, resulting in a sampling rate that is approximately $10^{24}$ times faster than brute-force simulations performed on supercomputers. These advancements represent a significant leap forward in the field of photonic quantum computing, opening the door to groundbreaking applications and advancing our comprehension of quantum computational potential.

The core concept involved the generation of photon pairs through spontaneous emission, resonating with a pump laser, to stimulate the emission of a second photon pair within a gain medium~\citep{jiuz25}. The experiment employed transform-limited laser pulses with a wavelength of 775 nm, focused on periodically poled potassium titanyl phosphate (PPKTP) crystals, to create two-mode squeezed states (TMSS). A concave mirror played a crucial role by reflecting both the pump laser and the collinear TMSS photons, acting as seeds for the second parametric process. To compensate for birefringence walk-off between horizontally and vertically polarized TMSS photons, a quarter-wave plate was utilized. Furthermore, the dispersion between the pump laser and the TMSS was mitigated by designing the PPKTP crystals to eliminate frequency correlation. Visual representations of the experimental setups are thoughtfully provided in Figure~\ref{fig:JiuzhangQC}.

In their research, the team achieved high collection efficiency (0.918 at a waist of 125 $\mu$m and 0.864 at 65 $\mu$m) along with simultaneous high photon purity (0.961 and 0.946) without the need for narrowband filtering. Their double-pass approach demonstrated scalability to higher orders, making it possible to generate quantum light sources with greater brightness, which are nearly optimal for a range of applications.

\subsection{Xanadu's Photonic QPUs}

\subsubsection{Overview}

Recently, the Canadian quantum computing company Xanadu~\citep{Xanadu} introduced its revolutionary photonic quantum computer, named "Borealis." Notably, Borealis demonstrated its quantum advantage by accomplishing a computational task in just 36 microseconds, a feat that would have taken a conventional supercomputer over 9,000 years to complete~\citep{Borealis}. These achievements underscore the potential of fully programmable quantum computers to achieve quantum advantage and surpass classical computing capabilities~\citep{Borealis}.

\subsubsection{The photonic quantum computer Borealis}

Xanadu's quantum technology primarily relies on Silicon quantum photonic chips~\citep{silicon}, forming the core of a room-temperature-operating quantum computer distinguished by its scalability. The realization of quantum computational advantage through Xanadu's photonic quantum computer, known as Borealis, is comprehensively examined in~\citep{Borealis}. 
In this context, Xanadu's Borealis stands out as a photonic processor renowned for its dynamic programmability across all implemented gates~\citep{Borealis3}, demonstrates quantum computational advantage~\citep{Borealis}.

The research employs Gaussian boson sampling (GBS)~\citep{GBS}, utilizing Borealis with 216 squeezed modes entangled through a three-dimensional connectivity configuration~\citep{Borealis5}, achieved through a time-multiplexed and photon-number-resolving architecture. Notably, Borealis completes this task within an astonishing 36 microseconds, while the most advanced classical algorithms and supercomputers would require over 9,000 years to accomplish the same feat~\citep{Borealis}. This performance improvement is remarkable, surpassing the advantage demonstrated by earlier photonic machines by a factor exceeding 50 million ~\citep{Borealis}. 

\begin{figure*}
    \centering
    \includegraphics[width=\textwidth]{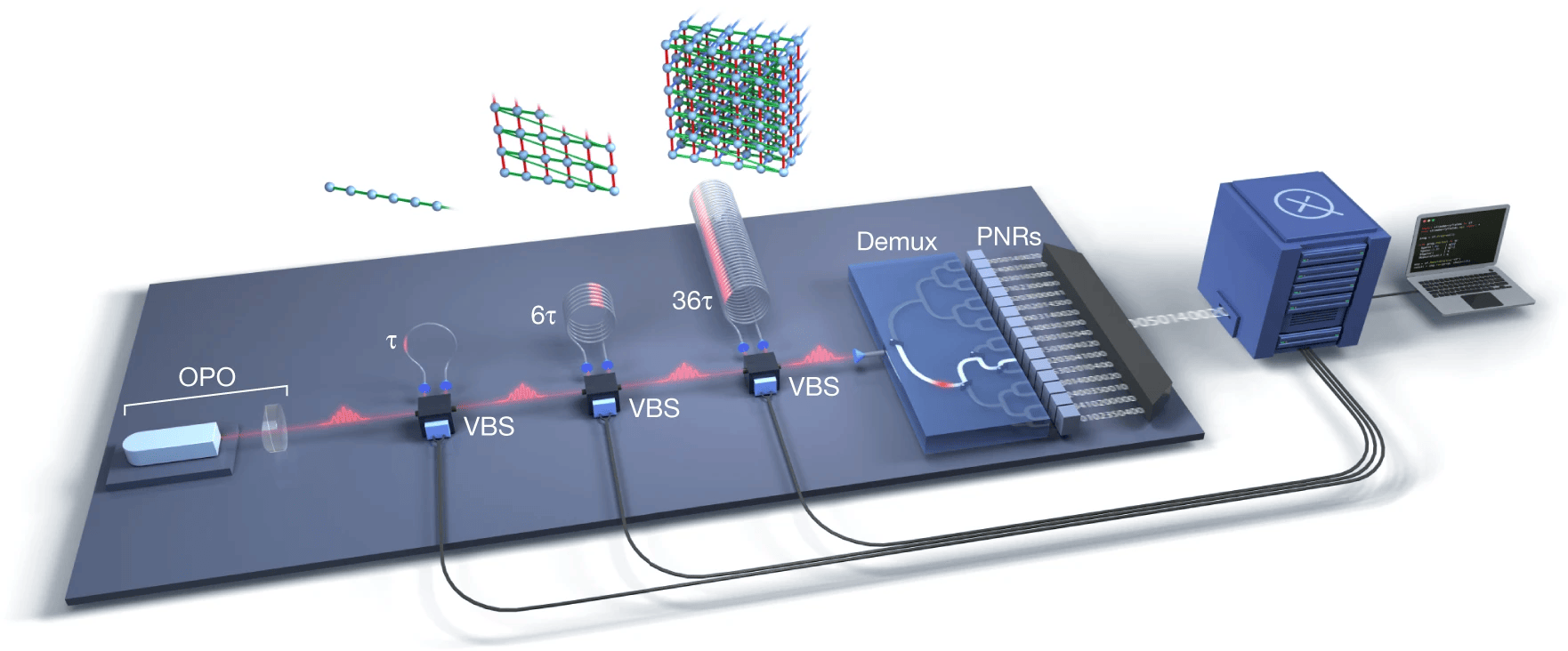}
    \caption{\textbf{High-dimensional GBS from a fully programmable photonic processor, Borealis:}
A  series of periodic pulses composed of single-mode squeezed states generated by a pulsed Optical Parametric Oscillator (OPO) is introduced into a sequence of three dynamically adjustable loop-based interferometers. Each of these loops consists of a Variable Beam Splitter (VBS) equipped with a programmable phase shifter, as well as an optical fiber delay line. The resulting Gaussian state at the interferometer's output is then directed towards a 1-to-16 binary switch tree (demux), which partially separates the output before it is measured by Photon Number Resolvers (PNRs). This recorded sequence comprises 216 photon numbers, and it is considered one sample, achieved in approximately 36 microseconds. The optical fiber delays, along with accompanying beamsplitters and phase shifters, facilitate the implementation of gates between both temporally adjacent and distant modes, thereby enabling high-dimensional connectivity within the quantum circuit. Additionally, above each stage of the loop, there is a visual representation of the multipartite entangled Gaussian state being progressively generated. In the first stage ($\tau$), two-mode programmable gates (depicted in green) are applied to adjacent modes in one dimension. In contrast, the second ($6\tau$) and third ($36\tau$) stages create connections between modes separated by six and 36 time bins in the second and third dimensions, respectively (illustrated by red and blue connections). Each operation of this device necessitates the specification of 1,296 real parameters, corresponding to the settings for all Variable Beam Splitter (VBS) units in the sequence. 
Reproduced under a Creative Commons Attribution 4.0 International License from Ref.~\citep{Borealis}.
}
    \label{fig:borealis}
\end{figure*}

\begin{figure*}
    \centering
 \includegraphics[width=\textwidth]{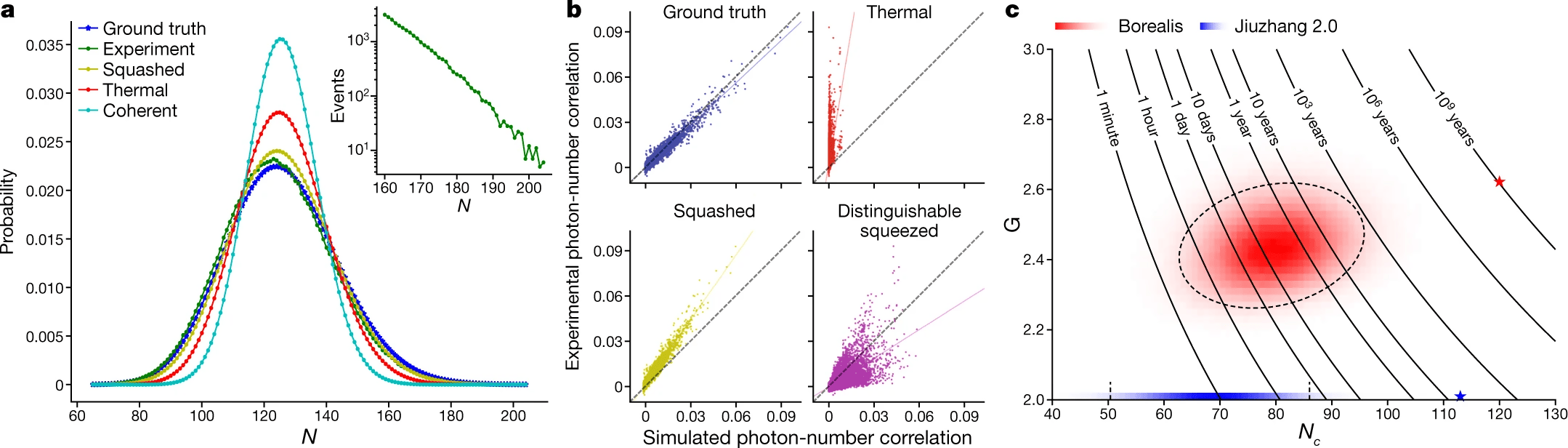}
    \caption{\textbf{A comparative analysis of the quantum computational advantages achieved by Borealis vs Jiuzhang 2.0 quantum computers.}
\textbf{a}) The measured photon statistics from $10^6$ samples of a high-dimensional Gaussian state are compared with numerical simulations based on various hypotheses. The inset displays the same distribution using a logarithmic scale, revealing substantial support for photon counts extending beyond 160 photons, up to 219.
\textbf{b}) A scatter plot illustrates the comparison between experimentally obtained two-mode cumulants $C_{ij}$ and those predicted by four different hypotheses. A perfect fit (depicted in plot) would result in experimentally obtained cumulants lying on a straight line at a 45-degree angle. Notably, only the ground truth hypothesis provides a good explanation for the cumulants. Furthermore, for a fair comparison, all hypotheses share identical first-order cumulants, i.e., the mean photon count in each mode.
\textbf{c}) The distribution of classical simulation times is depicted for each sample obtained in this experiment, denoted as "Borealis" in red, and for "Jiuzhang 2.0" in blue. For each sample from both experiments, a pair ($N_c, G$) is calculated, and a frequency histogram is constructed based on this two-dimensional space. It's important to note that all samples from Jiuzhang 2.0, being threshold samples, have $G = 2$, while samples from Borealis, which include collisions and photon-number resolution, have $G \ge 2$. The samples with the highest complexity in each experiment are marked with stars; these samples are found at the extreme end of their respective distributions and occur infrequently. The lines representing equal simulation times based on $N_c$ and G, are overlaid for each experiment. Dashed lines delineate boundaries corresponding to two standard deviations in the plotted distributions to aid visualization. 
Reproduced under a Creative Commons Attribution 4.0 International License from Ref.~\citep{Borealis}.
}
    \label{fig:comparison}
\end{figure*}

The study represents a substantial Gaussian boson sampling experiment, capturing events involving up to 219 photons with a mean photon number of 125. This significant achievement marks a pivotal step toward the realization of practical quantum computing and underscores the viability of photonics as a platform for advancing toward this goal. Figure~\ref{fig:borealis} present an illustration of the fully programmable photonic processor, Borealis. Furthermore, Figure~\ref{fig:comparison} provides a comparative analysis of the quantum computational capabilities achieved by Borealis alongside Jiuzhang 2.0 quantum computers.

\subsubsection{Xanadu's quantum cloud}

Accessible through Xanadu's Quantum Cloud (XQC) platform, Borealis provides direct access to their photonic Quantum Processing Units (QPUs), enabling remote utilization of their capabilities. Moreover, Xanadu's commitment to advancing quantum research is evident through their open-source library, PennyLane, which seamlessly integrates quantum machine learning, computing, and chemistry. PennyLane can interface with various quantum processors, including those offered by IBM~\citep{ibm} and Rigetti~\citep{rigetti0}. The successful realization of quantum computational advantage through Borealis' dynamic programmability underscores the promise of photonic quantum computing and contributes significantly to the advancement of quantum computing technology.

\section{Applied Quantum Computing}

Quantum computing is steadily advancing, catching the attention of governments and industry leaders. Its transformative potential spans various domains, promising swift and precise solutions that currently elude classical computers. Its impact encompasses pharmaceutical research, drug design and discovery~\citep{drugdiscovery}, financial modeling~\citep{Pricing100}, artificial intelligence and machine learning~\citep{MLappl}, combinatorial optimization~\citep{Ebadi}, logistics~\citep{cargoOPTO}, cybersecurity~\citep{crypsec}, cryptographic and encryption methodologies~\citep{Crpto1120K,Crpto16,Crpto24,Crpto6,Crpto23,Crpto10,Crpto9,Crpto8,Crpto7,Crpto25}. This relentless potential extends into fields such as electronic materials' discovery, computational chemistry, weather forecasting and climate change, manufacturing and industrial design, and more~\citep{NISQ000,NISQ001,NISQ002}. 
The horizons of quantum computing are marked by endless possibilities, and its ascent holds the promise of reshaping our understanding of what is achievable across numerous scientific, industrial, and technological frontiers~\citep{NISQ000,NISQ001,NISQ002}.

Herein, it is important to emphasize our deep appreciation for the dedicated research efforts that have contributed to the advancement of all disciplines within quantum computation and information processing. However, in this section, we specifically focus on a selection of cutting-edge quantum computing applications.

\subsection{Quantum Machine learning} 

Recently in~\citep{MLappl} a collaborative research effort involving scientists from prestigious institutions such as Harvard, Berkeley, Caltech, and Microsoft culminated in the project titled "Quantum Advantage in Learning from Experiments." Within this initiative, the researchers conducted experiments employing a superconducting quantum processor, up to 40 qubits. Their findings unequivocally demonstrated a significant quantum advantage, even in the presence of noise, within contemporary NISQ platforms~\citep{jpres18}.

The primary focus of their investigation was the exploration of how quantum technology could augment our capacity to unveil hitherto unknown natural phenomena. These experimental endeavors showcased the potential of supervised and unsupervised machine learning models~\citep{QML1,QML2} when fed with data derived from quantum-augmented experiments utilizing quantum memory and quantum processing. Remarkably, these models exhibited the capability to predict properties and unveil underlying structures within physical systems that lay beyond the purview of traditional experimentation.

Crucially, their research substantiated that quantum computing systems can acquire knowledge from a notably reduced number of experiments in comparison to conventional methodologies, a phenomenon that holds true across various tasks. This quantum advantage is most pronounced in the prediction of physical system properties, the execution of quantum principal component analysis on noisy states, and the approximation of models representing physical dynamics, all achievable with a significantly reduced experimental load.

Notably, while previous investigations into quantum advantage predominantly centered on computational tasks with predefined inputs, the researchers deviated from this paradigm by concentrating on learning tasks aimed at gaining insights into previously unknown physical systems. This novel approach contributes fresh perspectives toward comprehending and attaining quantum advantage within the domains of quantum machine learning~\citep{QML2,QML3} and quantum sensing~\citep{quasensing}. 
In the academic literature, an extensive body of dedicated research efforts~\citep{NISQ295,NISQ296,NISQ297,NISQ299,NISQ298,NISQ301,NISQ303,NISQ322,NISQ307,NISQ304,NISQ302,NISQ330,NISQ305,NISQ292,NISQ294,NISQ326,NISQ331,NISQ308,NISQ293,NISQ323,NISQ310,NISQ320,NISQ329,NISQ313,NISQ312,NISQ314,NISQ316,NISQ311,NISQ317,NISQ306,NISQ319,NISQ321,NISQ318,NISQ324,NISQ332,NISQ325,NISQ328} has made substantial contributions to the progression of Quantum Machine Learning (QML) and Artificial Intelligence (AI).

\subsection{Quantum Simulation/Material Science}

Quantum computing holds the potential to simulate complex quantum systems, offering valuable insights into the behavior of molecules and materials. This has profound implications for fields such as materials science and drug discovery. Traditional classical computers face limitations in accurately modeling intricate molecular interactions. In~\citep{QSimuIBM1,QSimuIBM,QSimuIBM4}, a notable application involves the accelerated development of chemical products, including catalysts and surfactants, within the chemical and petroleum industries. Quantum computers are being employed by companies in these sectors, such as IBM, to model substances like beryllium hydride (BeH$_2$) and lithium hydride (LiH). These quantum-powered simulations pave the way for the creation of new catalysts aimed at emissions reduction and enhanced surfactants to improve subsurface recovery~\citep{QSimuIBM1,QSimuIBM,QSimuIBM4,ExxonMobil}.

In the field of chemistry, quantum computers offer the promising potential to enhance the precision of direct molecule simulations. This advancement holds the prospect of enabling a more accurate investigation into critical reactions and systems that currently remain beyond the reach of classical computing methods. Initial assessments suggest that, in the fault-tolerant era, quantum computers will eventually have the capacity to address such complex challenges~\citep{NISQ253,NISQ252,NISQ251}.

There have been proposed strategies for leveraging quantum computers to solve models involving strongly correlated electrons~\citep{NISQ250}.  Among the molecules garnering significant attention is FeMeCo~\citep{NISQ249,NISQ248}. Preliminary applications have utilized Variational Quantum Eigensolver (VQE) techniques to estimate the energy levels of water molecules~\citep{NISQ257} and small hydrogen chains~\citep{NISQ166,NISQ157}. For a more exploration of the applicability of VQE to chemical problems, detailed studies can be found here~\citep{NISQ258}. Additionally, Variational Quantum Algorithms (VQAs) have been proposed to simulate exciton dynamics within molecules~\citep{NISQ264,NISQ263}. For a deeper understanding of the application of quantum computers in  chemistry, comprehensive review papers are available in~\citep{NISQ266,NISQ265}.

Moreover, within the realm of high-energy physics, initial proof-of-principle simulations have shown promise in studying the temporal dynamics of a simplified Quantum Chromodynamics (QCD) model. These simulations have been conducted using both trapped ions~\citep{NISQ271} and superconducting circuits~\citep{NISQ134}. Variational Quantum Simulators (VQS) have also been harnessed for such simulations, achieving success with system sizes of up to 20 qubits and 15 variational parameters~\citep{NISQ272}. Quantum neural networks have been proposed as tools to assist in the analysis of data generated by experiments in high-energy physics~\citep{NISQ282,NISQ281}. For comprehensive insights into this field, valuable review papers are available~\citep{NISQ284,NISQ283}.

Quantum simulation finds intriguing applications in diverse domains, including nuclear physics~\citep{NISQ290,NISQ289}, quantum control~\citep{NISQ288}, and material design~\citep{NISQ287}, among others. For a comprehensive overview of quantum simulators, readers are referred to~\citep{NISQ267,NISQ291}.

\subsection{Quantum Natural Language Processing}

Cambridge Quantum Computing (CQC) has recently unveiled an outstanding advancement~\citep{lambeqCQC,lambeqHPC} in the field of Natural Language Processing (NLP) by introducing the pioneering quantum-based natural language processing (QNLP) toolkit and library~\citep{lambeqarXiv,lambeqGithub}. 
This innovative toolkit, known as "lambeq,"~\citep{lambeqarXiv} represents a seminal achievement as the world's inaugural software toolkit capable of translating sentences into quantum circuits. Its primary objective is to expedite the development of pragmatic QNLP  with real-world relevance~\citep{lambeqAPP4}. These applications span a spectrum of domains, encompassing automated dialogue systems, text mining, language translation, text-to-speech synthesis, language generation, and bioinformatics~\citep{lambeqAPP2,lambeqAPP3,lambeqAPP4}. 

\begin{figure*}
    \centering
    \includegraphics[width=0.9\textwidth]{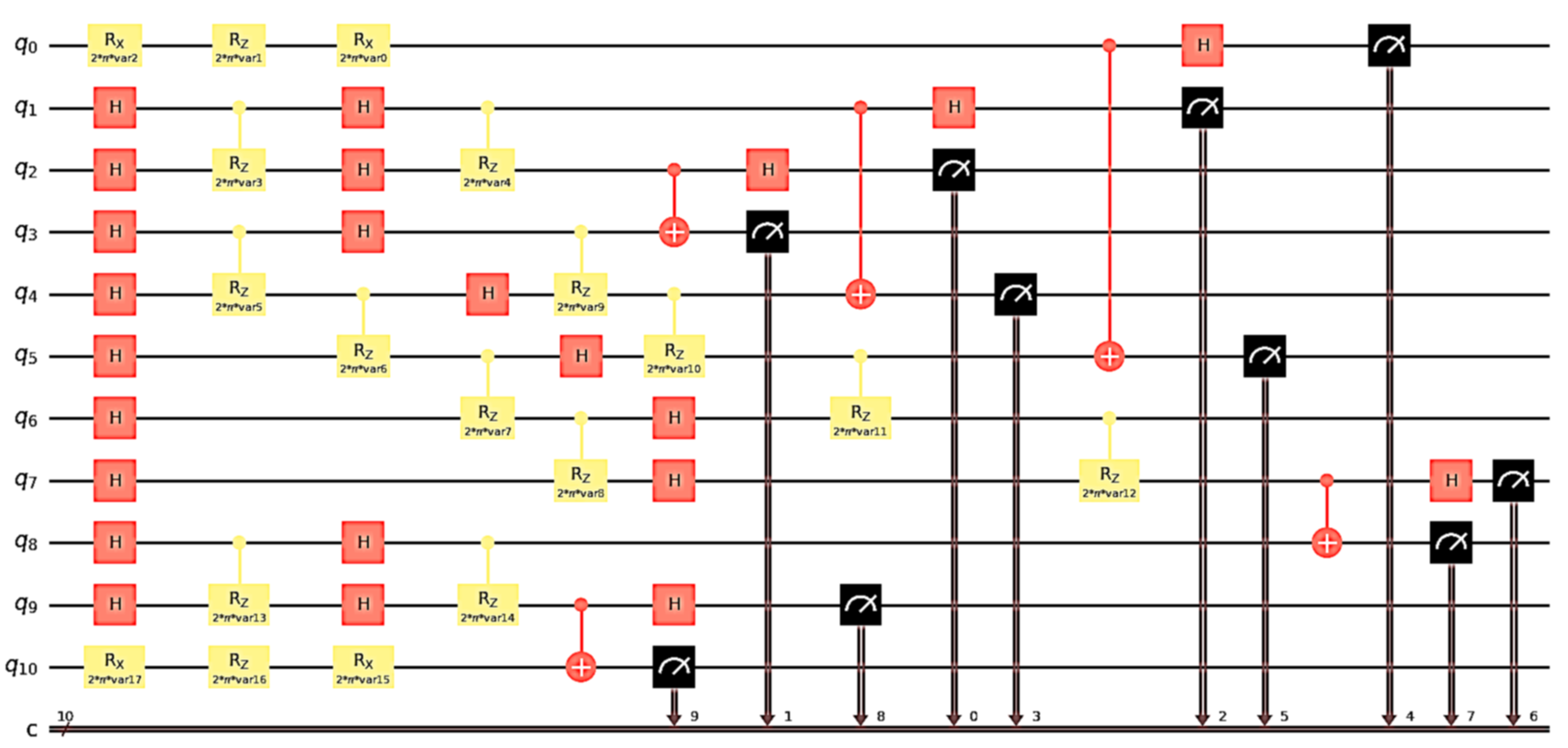}
    \caption{A ten-qubit quantum circuit representation that corresponds to the sentence “John walks in the park” in IBM's Qiskit format, with H stands for Hadamard gate and $R_Z$ denotes the rotation gates. Reproduced with formatting alterations under a Creative Commons license from Ref.~\citep{lambeqarXiv}.}
    \label{fig:lambeq}
\end{figure*}

Figure~\ref{fig:lambeq} presents a quantum circuit that corresponds to the text "John walks in the park"~\citep{lambeqarXiv}, affording readers a visual depiction of the toolkit's operational prowess. 
For an in-depth technical exposition, this article~\citep{lambeqarXiv} furnishes a comprehensive exposition detailing the fundamental architecture of lambeq. Moreover, it offers a detailed explication of the toolkit's essential modules, highlighting their significance. Further insights into lambeq's implementation details are explained in~\citep{lambeqAPP2,lambeqAPP3,lambeqAPP4}.

\subsection{Quantum Computing in Finance}

Quantum computing companies Pasqal~\citep{Pasqal} and Multiverse~\citep{Multiverse} Computing, in collaboration with Crédit Agricole~\citep{Agricole}, unveiled~\citep{HPCfinance} the successful conclusion of a 1.5-year Proof of Concept (POC) study~\citep{Pricing}. This study aimed to assess the effectiveness of an algorithmic approach inspired by quantum computing, as well as the potential applications of quantum computers in two critical areas: the valuation of financial instruments and the evaluation of credit risks~\citep{Pricing,RiskManag}. The outcomes of their research highlight the development of the inaugural quantum-enhanced machine learning (ML) algorithm designed for forecasting credit rating downgrades. These POCs~\citep{Pricing,RiskManag,tensorPDEs} serve as tangible demonstrations of quantum computing's prospective and real-world utility within the realm of finance.

\subsection{Solving Hard Graph Problems}

Combinatorial optimization, a prevalent challenge across scientific and technological fields, often involving problems that exhibit inherent computational complexity, serving as a cornerstone for understanding contemporary computer science complexity classes~\citep{Ebadi1}. The utilization of quantum machines to expedite the solution of such problems has been explored theoretically for over two decades through various quantum algorithms~\citep{Ebadi3,Ebadi2,Ebadi4}. Typically, a relevant cost function is encoded within a quantum Hamiltonian~\citep{Ebadi5}, and the goal is to identify its lowest-energy state, initiated from a generic initial state. This is achieved either through adiabatic evolution~\citep{Ebadi2} or a variational approach~\citep{Ebadi3}, involving iterative optimization processes~\citep{Ebadi7,Ebadi6}. The computational performance of these algorithms has been examined both theoretically~\citep{Ebadi4,Ebadi8,Ebadi10,Ebadi9,Ebadi11,Ebadi13,Ebadi12} and experimentally~\citep{Ebadi14,Ebadi16,Ebadi15}, mostly in small quantum systems with shallow quantum circuits or systems lacking the presumed many-body coherence essential for achieving quantum advantage~\citep{Ebadi17,Ebadi18}. However, these investigations offer limited insights into algorithm performance in the most interesting scenario, characterized by large system sizes and deep circuitry~\citep{Ebadi20,Ebadi19}.

Achieving quantum acceleration in addressing computationally challenging dilemmas stands as the primary imperative within the realm of quantum information science. 
\textit{Ebadi et al.}~\citep{Ebadi} conducted a pioneering investigation employing Rydberg atom arrays, a neutral atom quantum processor comprising up to 289 coupled qubits in two spatial dimensions, to explore quantum optimization algorithms for tackling practical dilemmas, specifically the maximum independent set (MIS) problem. The MIS problem serves as a quintessential example of a combinatorial optimization problem that falls into the category of nondeterministic polynomial time–complex challenges.

In prior research, proposals had been advanced for the efficient encoding of specific challenging combinatorial optimization problems within neutral-atom quantum processors. However, in this groundbreaking publication~\citep{Ebadi}, the authors not only executed the inaugural instance of effective quantum optimization on an operational quantum computing platform but also demonstrated an unprecedented level of quantum hardware capabilities. These findings signify the initial stride towards harnessing valuable quantum advantages in addressing demanding optimization issues that hold relevance across various industrial sectors.

Additionally, one prominent problem  in this realm is the NP-complete Max-Cut problem~\citep{NISQ353}, which has garnered substantial  attention in the quantum computing community. The Quantum Approximate Optimization Algorithm (QAOA) was introduced for solving the Max-Cut problem and has been demonstrated on instances involving up to 19 qubits~\citep{NISQ168}. Subsequent research has extended the QAOA's applicability to the Max-Cut problem ~\citep{NISQ357,NISQ356,NISQ359,NISQ358}. 

Quantum computers have also been investigated for tackling other combinatorial optimization problems, including the knapsack problem ~\citep{NISQ363}, protein folding~\citep{NISQ361,NISQ360,NISQ362}, the k-clique problem~\citep{NISQ365}, and graph coloring~\citep{NISQ364}.

\subsection{Logistics Optimization}

Quantum computing, a game-changing technology, is finding applications in logistics and manufacturing, promising to revolutionize optimization tasks. Quantum-South, a  quantum technology company, exemplifies this potential. 
In June 2022, Quantum-South showcased their breakthrough in cargo loading optimization using quantum annealers through Amazon Braket~\citep{AWS}. This approach improves cargo plans, potentially streamlining the allocation of passengers, weight, and volume on flights, boosting operational efficiency and reducing analysis time~\citep{cargo}. 

Quantum-South's pioneering work has not gone unnoticed in the industry. IAG Cargo~\citep{IAGCargo}, a division of International Airlines Group, trialed Quantum-South's quantum algorithms for air cargo optimization, highlighting growing industry interest. Additionally, Quantum-South explored ULD flight loading optimization through quantum computing, customizing algorithms to streamline the loading process in partnership with IAG Cargo~\citep{cargo,ULD,cargoOPTO,cargoOPTO1}. These developments underscore the transformative impact of quantum computing in logistics and manufacturing.

\subsection{Quantum Numerical Solvers}

Quantum computing has demonstrated its theoretical capability to achieve remarkable speed improvements in specific numerical problem-solving tasks~\citep{NISQ333}. Notably, the HHL algorithm and subsequent enhancements promise an exponential acceleration in solving linear systems 
compared to classical computer solvers~\citep{NISQ335,NISQ334}. 
In addition to these advancements, alternative approaches utilizing Variational Quantum Algorithms (VQAs) have been proposed to address linear system problems~\citep{NISQ337,NISQ336,NISQ338}. Furthermore, quantum-assisted methods have exhibited proficiency in solving classically solvable systems, including those with dimensions as large as $2^{300} \times 2^{300}$~\citep{NISQ339}.

Classical computers often require substantial computational resources to solve non-linear differential equations~\citep{,NISQ340}. However, proof of concept results from experiments conducted on an IBM Quantum device suggest that variational quantum computing can efficiently tackle such non-linear problems, exemplified by the nonlinear Schrödinger equation~\citep{NISQ341}. 
Moreover, the scientific community is exploring differentiable quantum circuits as a means to address complex problems ~\citep{NISQ343}, with considerable interest directed towards employing Near-Term Quantum (NISQ) algorithms for solving intricate equations like the Navier-Stokes equations~\citep{NISQ345,NISQ344}. 
Quantum computers exhibit potential in several other numerical problem-solving applications, as evidenced by a body of literature encompassing references~\citep{NISQ12,NISQ346,NISQ347,NISQ340,NISQ348,NISQ341,NISQ344,NISQ343,NISQ350,NISQ342,NISQ345}.

The horizons of quantum computing are marked by endless possibilities, and its ascent holds the promise of reshaping our understanding of what is achievable across numerous scientific, industrial, and technological frontiers.

\section{Conclusion and Outlook}\label{conc}

NISQ quantum computing has set the stage for a technological revolution, 
garnering significant attention and investment. This is due to its potential to tackle complex problems--more 
effectively than conventional computing schemes--with far-reaching implications across various sectors, from scientific breakthroughs to national security. This has prompted research agencies and tech giants to establish quantum computing departments. 
However, as we stand at the intersection of quantum dreams and real-world challenges, it is essential to acknowledge that errors, challenges in experimental control techniques, and the limited  number of qubits remain substantial obstacles that hinder the realization of the full potential of current Noisy Intermediate-Scale Quantum computers.

The road to fault-tolerant quantum (FTQ) computers, 
which are quantum computers capable of handling these intricate challenges, necessitates substantial advancements in refining experimental control techniques for qubit gate operations, qubit quality, error correction, and a significant increase in the number of available qubits. 
While recent experimental progress has sparked optimism, building a fault-tolerant quantum computer is a demanding endeavor, likely requiring at least another decade of tireless development.

In the era of rapid technological advancement, quantum computing appears to defy the trend by progressing more gradually. 
The presented proof-of-concept 
cases underscore the immense potential of quantum computing and quantum processors. Despite recent strides in quantum hardware and algorithms, practical applications remain limited. However, as quantum computing technologies continue to advance, we anticipate the emergence of more practical use cases. We should closely monitor developments in academia and eagerly await well-researched applications from leading organizations worldwide. These applications may soon bring quantum computing into our lives in transformative and positive ways. 

The journey toward unlocking the full potential of quantum computing is ongoing, and its ultimate impact remains a promising prospect on the horizon. As we bid farewell to NISQ era and delve deeper into the FTQ realm, we find ourselves at the doorstep of an exceptional future, where  things that seem impossible today may soon become not only possible but routine.

\begin{acknowledgments}

The views and conclusions expressed in this work are solely those of the authors and do not reflect the official policy or position of Google's Quantum AI, UST China, Zuchongzhi Quantum Computers, Jiuzhang Quantum Computers, Xanadu's Photonic QPUs, Cambridge Quantum Computing, Pasqal, Multiverse Computing, Quantum-South, IAG Cargo, Microsoft, Crédit Agricole, Harvard University, MIT, or any affiliated organizations. 
“This research received no specific grant from any funding agency in the public, commercial, or not-for-profit sectors.” 

\end{acknowledgments}

\section*{Author contributions}

MA Conceptualization, Methodology, Resources, Data curation, Visualization, Investigation, Validation, Writing - Original Draft, Review and Editing. HE Investigation, Validation, Reviewing and Editing.  All authors have approved the final manuscript.  


\section*{Data Availability Statement}
The datasets generated during and/or analyzed during the current study are included within this article.

\end{document}